\documentclass[twoside]{LCWS11}
\usepackage[latin1]{inputenc}
\usepackage[dvips]{graphicx,epsfig,color}
\usepackage{wrapfig,rotating}
\usepackage{slashed}
\usepackage{amssymb,amsmath,array}
\usepackage{cite}

\newcommand{\be}{\begin{equation}}
\newcommand{\ee}{\end{equation}}
\newcommand{\bea}{\begin{eqnarray}}
\newcommand{\eea}{\end{eqnarray}}

\newcommand{\bi}{\begin{itemize}}
\newcommand{\ei}{\end{itemize}}

\newcommand\T{\rule{0pt}{2.5 ex}}

\newcommand\B{\rule[-1.5ex]{0pt}{0pt}}

\begin{document}
\title{
Chargino Production at a future LC in the MSSM with complex Parameters: NLO Corrections} 
\author{Aoife Bharucha
\vspace{.3cm}\\
II. Institut f\"{u}r Theoretische Physik, University of Hamburg, Luruper Chaussee 149,\\ D-22761 Hamburg, Germany
}

\maketitle

\begin{abstract}
We calculate chargino production at a future linear collider, including full one-loop contributions in the MSSM with complex parameters. To achieve this, we require a comprehensive approach to renormalisation in the chargino and neutralino sector, for the case where parameters can be complex. Using the field renormalisation as developed in Refs.~\cite{Fowler:2009ay,AlisonsThesis,bfmw}, we investigate the parameter renormalisation in the complex case. We also study the numerical role of the choice of renormalisation scheme as in Refs.~\cite{AlisonsThesis,bfmw}. Finally we present new results showing the effect of the phases on the resulting cross section at NLO.

\end{abstract}

\section{Introduction}

In light of latest LHC data~\cite{Aad:2011ib,Chatrchyan:2011zy} and fits to it (see e.g. Ref~\cite{Buchmueller:2011sw}), it is thought that the charginos and neutralinos could be among the lightest supersymmetric particles, and therefore within reach of a linear collider. Their production should allow a leading order determination of $M_1$, $M_2$, $\mu$ and $\tan\beta$, at the percent level (e.g. Ref.~\cite{Desch:2003vw}), as well as access to phases arising at next-to-leading order (NLO) via asymmetries (e.g. Ref.~\cite{Rolbiecki:2007se}).
Due to the high sensitivity of measurements in the clean LC environment, higher order corrections would be crucial. The on-shell renormalisation of the chargino-neutralino sector was investigated for the minimal supersymmetric standard model (MSSM) with real parameters in Refs.~\cite{Eberl:2001eu,Fritzsche:2002bi,Oller:2003ge,Oller:2005xg,Fowler:2009ay,Chatterjee:2011wc}. Working in the on-shell scheme for the complex MSSM requires a consistently defined framework. When carrying out the field renormalisation, absorptive parts of loop integrals must be included~\cite{Fowler:2009ay}. For the parameter renormalisation, the masses required on-shell must be carefully chosen~\cite{AlisonsThesis,Chatterjee:2011wc}, and in addition the phases renormalisation must be considered~\cite{AlisonsThesis,bfmw}. 
In the following we will introduce the complex MSSM and our approach to its renormalisation, as described in Ref.~\cite{bfmw}. We further investigate the numerical role played by the choice of renormalisation. Finally we present results for the dependence of the NLO corrections to the cross-section for chargino production on the complex phases of MSSM parameters.


\section{Renormalization of the chargino and neutralino sector of the MSSM}
\label{sec:3}
Charginos and neutralinos are the mass eigenstates of gauginos and higgsinos, and the mass matrix for the charginos is given by 
\begin{equation}\label{eq:X}
X=
\left( \begin{array}{cc}
M_2 & \sqrt{2} M_W s_\beta  \\
\sqrt{2} M_W c_\beta  & \mu
\end{array} \right),
\end{equation}
where $s_\beta/c_\beta\equiv\sin\beta/\cos\beta$. The chargino masses are obtained by diagonalising the matrix via the bi-unitary transformation $\displaystyle\mathbf{M}_{\tilde{\chi}^+}=U^* X V^\dag$.
The mass matrix for the neutralinos in the $(\tilde{B},\tilde{W},\tilde{H}_1,\tilde{H}_2)$ basis is given by
\begin{equation}\label{eq:Y}
Y =\left( \begin{array}{cccc}
M_1 & 0 & -M_Z c_\beta s_W & M_Z s_\beta s_W \\
0   & M_2 & M_Z c_\beta c_W & -M_Z s_\beta c_W \\
-M_Z c_\beta s_W & M_Z c_\beta c_W & 0 & -\mu \\
M_Z s_\beta s_W & -M_Z s_\beta c_W & -\mu & 0 \end{array} \right).
\end{equation}
Since $Y$ is complex symmetric, its diagonalisation requires only one unitary matrix $N$, via $\mathbf{M}_{\tilde{\chi}^0}=N^*Y N^\dag$, giving us the neutralino masses.
Clearly the additional parameters that enter this sector are $M_1$, $M_2$ and $\mu$.

We renormalise the chargino and neutralino fields in the most general way, making use of separate renormalisation constants (RCs) for the incoming and outgoing fields, i.e. coefficients $\delta Z^{L/R}_{\pm,ij}$ and $\bar{Z}^{L/R}_{\pm,ij}$ respectively for left and right-handed charginos, and $\delta Z^{L/R}_{0,ij}$ and $\bar{Z}^{L/R}_{0,ij}$ for left and right-handed neutralinos, as defined in Ref.~\cite{Fowler:2009ay,AlisonsThesis,bfmw}.
Renormalising the matrices $X$ and $Y$ in the case where $M_1$ and $\mu$ may be complex quantities, the five independent parameters $\delta|M_1|$, $\delta|M_2|$, $\delta|\mu|$, $\delta\phi_{M_1}$ and $\delta\phi_\mu$ are renormalised via
\begin{align}
 |M_1|\rightarrow|M_1|+\delta|M_1|,\;\;\;\;\ |M_2|&\rightarrow|M_2|+\delta|M_2|,\;\;\;\;\|\mu|\rightarrow|\mu|+\delta|\mu|, \nonumber\\
\phi_{M_1}\rightarrow \phi_{M_1}+\delta\phi_{M_1},&\;\;\;\;\phi_{\mu}\rightarrow \phi_{\mu}+\delta\phi_{\mu}\label{deltamu}.
\end{align} 
For consistency and to ensure the gauge-boson masses are defined on-shell throughout, $\tan\beta$, $M_W$, $M_Z$ and $\sin\theta_W$ are renormalised as defined in Ref.~\cite{bfmw}. Our procedure closely follows Refs.~\cite{Fritzsche:2002bi,Fritzsche:2005,Fowler:2009ay,AlisonsThesis}, and differs from the method used in Ref.~\cite{Oller:2005xg}, where the mixing matrices are renormalised using the proposal of Ref.~\cite{Kniehl:1996bd}, which draws a parallel with the renormalisation of the CKM matrix, discussed in more detail in Sec.~\ref{sec:3.2}.

\subsection{Field renormalisation}
\label{sec:3.2}
In order to obtain expressions for the field RCs, we impose the standard on-shell conditions as defined in Ref.~\cite{bfmw}. Adopting the most general approach, we do not impose the hermiticity relation $\displaystyle \delta Z^{L/R}_{\pm,ij}\neq\delta\bar{Z}^{R/L}_{\pm,ji}$, and but instead use the additional conditions that the renormalised propagators retain the same Lorentz structure as the tree level propagators.
The expressions we find for the RCs~\cite{bfmw} obey the above hermiticity condition up to the absorptive parts of the loops integrals. Dropping these absorptive parts is possible in the real case up to the one-loop level, but when complex parameters are present products of these and the absorptive parts means that the on-shell conditions would no longer be satisfied~\cite{Fowler:2009ay,AlisonsThesis,bfmw}. We later investigate the size of the effect of ignoring these absorptive parts on predictions for physical observables.

\subsection{Parameter renormalisation}
For the parameters $|M_1|$, $|M_2|$ and $|\mu|$, on-shell conditions can easily be obtained by considering the case where $\phi_{M_1}$ and $\phi_\mu$ vanish.
\begin{table}
\begin{center}
\begin{footnotesize}
\begin{tabular}{c||c|c|c|c|c||c|c}
\hline
\T
\B
 & NNN & NNC & NCC & NCCb& NCCc&NCCb* & NCCc*\\
\hline
\hline
\T $\delta |M_1|$ & -1.468 & -1.465 &  -1.468 & 2517 & -3685&-365.4& -4.671\\
 $\delta |M_2|$ & -9.265 & -9.265 &  -9.410 & -9.410 & -9.410 &13.23 &13.23\\
 $\delta |\mu|$ & -18.48 & -18.98 &  -18.98 & -18.98 & -18.98 & -5.333 & -5.333\\
 $\Delta m_{\tilde{\chi}^0_1}$ & 0 & 0 & 0 & 2517. & -3681&-5.809&-0.522\\
 $\Delta m_{\tilde{\chi}^0_2}$ & 0& 0& -0.1446 & 0& 0.3560 &0& -0.4806\\
 $\Delta m_{\tilde{\chi}^0_3}$& 0 & -0.5018 &  -0.5016 & -0.8447 & 0& -354.9& 0\\
 $\Delta m_{\tilde{\chi}^0_4}$ & 0.3238 & -0.1775 & -0.1775 & 0.6851& -1.439& -0.1734&-0.1548\\
 $\Delta m_{\tilde{\chi}^{\pm}_1}$ & 0.1446 & 0.1445 &  0 & 0& 0&0&0\\
\B $\Delta m_{\tilde{\chi}^{\pm}_2}$ & 0.5012 & 0 &  0 & 0& 0 &0&0\\
\hline
\end{tabular} 
\caption{Finite parts of parameter RCs and mass corrections in $\mathrm{GeV}$ for the CPX scenario (see text). The columns denoted with an asterisk are for a higgsino-like scenario.}
\label{tab:RenormScheme}\
\end{footnotesize}
\end{center}
\end{table}
Expressions for $\delta|M_1|$, $\delta|M_2|$, $\delta|\mu|$ are derived by choosing three out of the total of six physical masses to be on-shell~\cite{AlisonsThesis,bfmw}. There are clearly three possibilities, three neutralinos $\chi^0_1, \chi^0_2,\chi^0_3$ (NNN), two neutralinos and one chargino $\chi^0_1, \chi^0_2,\chi^\pm_2$ (NNC) or one neutralino and two charginos $\chi^0_{1(2/3)}, \chi^\pm_1,\chi^\pm_2$ (NCC(b/c)). The difference between these was investigated in detail in Ref.~\cite{AlisonsThesis} for the case of the CPX scenario ($M_1=(5/3) (s_W^2/c_W^2) M_2$, $M_{\rm SUSY}=$500 GeV, $A_{q,l}$=900 GeV, $\phi_{M_1}$=0, $\phi_{\mu}$=0, $\phi_{M_3}=\pi/2$, $\phi_{A_{f3}}=\pi/2$,  $\phi_{A_{f1,2}}=\pi$, $\tan\beta=5.5$ and $M_{H^{\pm}}=132.1\,\mathrm{GeV}$, $M_2$=200 GeV and $\mu$=2000 GeV) and a higgsino-like variant of the CPX scenario ($\mu=200\,\mathrm{GeV}$ and $M_2=1000\,\mathrm{GeV}$).
As seen in Tab.~\ref{tab:RenormScheme}, not only should the choice be made such that external particles are on-shell, but also such that the given MSSM scenario is considered. 
To be more precise, in a gaugino-like scenario, where $M_1<M_2\ll\mu$, the values of the parameters $M_1$, $M_2$ and $\mu$ broadly determine the values of the masses of $\chi^0_1$, $\chi^0_2$/$\chi^\pm_1$ and $\chi^0_{3/4}$/$\chi^\pm_2$ respectively, such that NNN, NNC and NCC are suitable.
On the other hand a higgsino-like scenario, where $\mu\ll M_1<M_2$, would mean the values of the parameters $M_1$, $M_2$ and $\mu$ are closely related the values of the masses $\chi^0_3$, $\chi^0_4$/$\chi^\pm_2$ and $\chi^0_{1/2}$/$\chi^\pm_1$ respectively, such that NCCc is suitable.
Whichever scenario is being studied, one physical mass which can constrain each fundamental parameter must be chosen on-shell.
Failing to do so will result in numerically unstable RCs, therefore possibly taking large unphysical values.
This was also investigated in detail, including the case of strongly mixed scenarios, in Ref.~\cite{Chatterjee:2011wc}.

Choosing on-shell conditions suited to determine the phase RCs is not obvious. However, assuming an on-shell scheme as well as the requirement~\cite{AlisonsThesis},
 \begin{eqnarray}
\delta Z^R_{0,11}= \delta \bar{Z}^R_{0,11},\;\;\;\delta Z^L_{0,11}= \delta \bar{Z}^L_{0,11},\;\;\;\delta Z^R_{\pm,22}= \delta \bar{Z}^R_{-,22},\;\;\;\delta Z^L_{\pm,22}= \delta \bar{Z}^L_{-,22},
\end{eqnarray}
the expressions obtained for $\delta\phi_{M_1}$ and $\delta\phi_{\mu}$ were shown to be UV-convergent. Note that this conditions can be understood as imposing that the imaginary parts of the diagonal field RCs for the on-shell particles vanish. We therefore choose not to renormalise the phases, which is more convenient than a specific renormalisation scheme as the phases remain at their tree-level value.

\section{Numerical analysis}
We now wish to utilise the above renormalisation framework to calculate $\sigma(e^+e^-\to\tilde{\chi}^+_i\tilde{\chi}^-_j)$ at NLO, including full MSSM corrections and allowing parameters to be complex. The NLO corrections were calculated in the real case in Refs.~\cite{Oller:2003ge,Fritzsche:2005}. The details of our calculation and definitions for the parameters can be found in Ref.~\cite{bfmw}. Here we simply state the results for the scenario given in
Tab.~\ref{tab:Params}, varying the phases $\phi_{A_t}$, $\phi_{A_b}$, $\phi_{A_\tau}$, 
$\phi_{M_1}$, $\phi_{M_3}$ and $\phi_\mu$ within bounds taking into account the EDM constraints~\cite{Lee:2003nta}.

\begin{wraptable}{l}{.5\columnwidth}
\begin{footnotesize}
\centerline{\begin{tabular}{c|c||c|c}
\hline
 Par. & Value & Par. & Value\\
\hline
\hline
$|M_1|$ & 100 GeV& $M_2$ & 200 GeV\\
$|\mu|$ & 420 GeV & $M_{H^+}$ & 800 GeV\\
$|M_3|$ & 1000 GeV & $\tan\beta$ & 20\\
$M_{\tilde{q}_{1,2}}$ & 1000 GeV & $M_{\tilde q_3}$ & 500-800 GeV\\ 
$M_{\tilde{l}_{1,2}}$ & 400 GeV & $M_{\tilde l_3}$ & 500 GeV\\ 
$|A_q|$ & 1300 GeV & $|A_l|$ & 1000 GeV\\
\hline
\end{tabular}}
\end{footnotesize}
\caption{Table of parameters, where $A_{q/l}$ are the quark/lepton trilinear couplings.}
\label{tab:Params}
 \end{wraptable}

\begin{figure}
\begin{center}
\includegraphics[scale=.45]{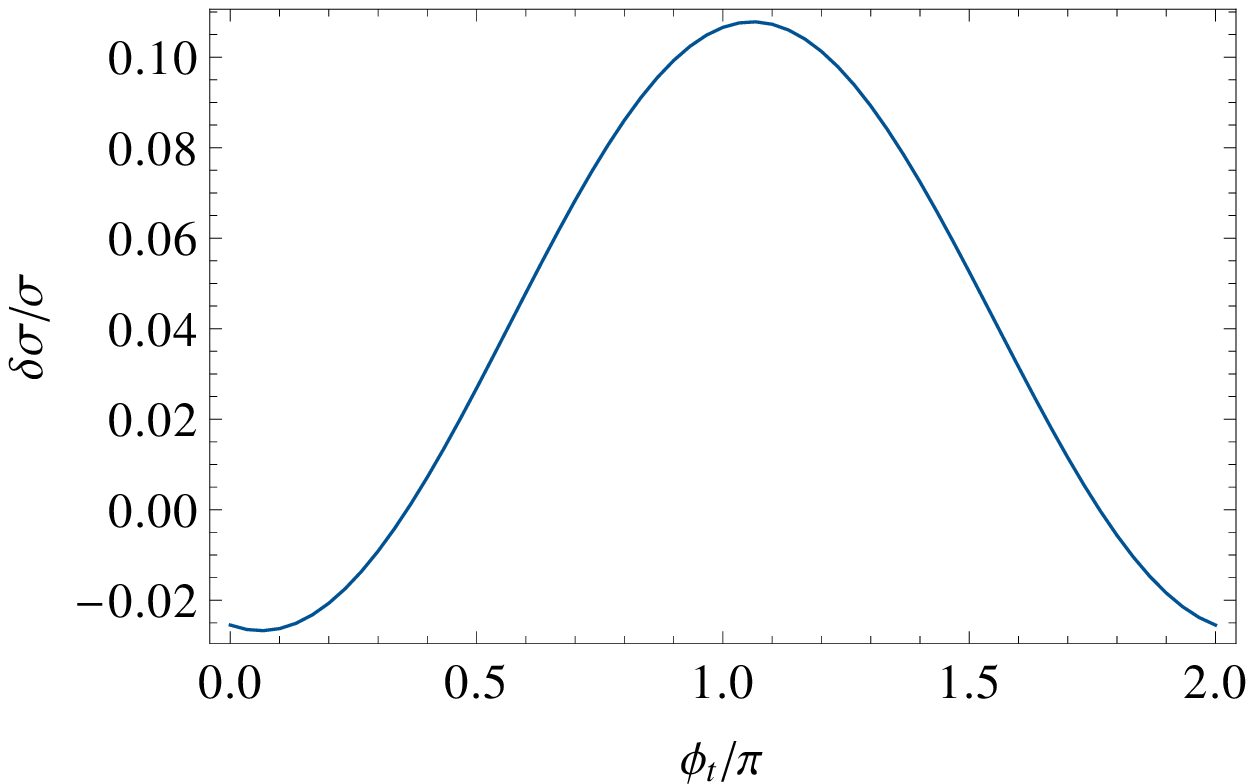}
\includegraphics[scale=.45]{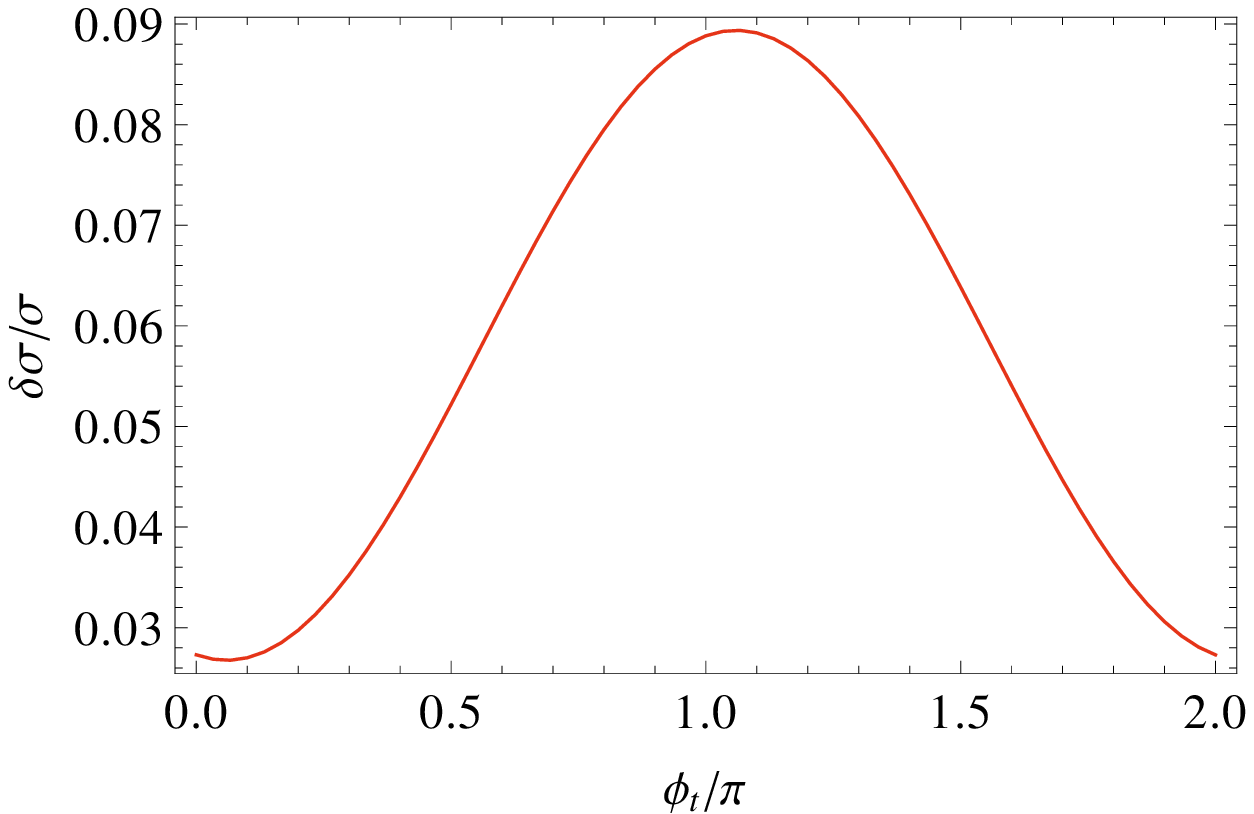}\\
\includegraphics[scale=.45]{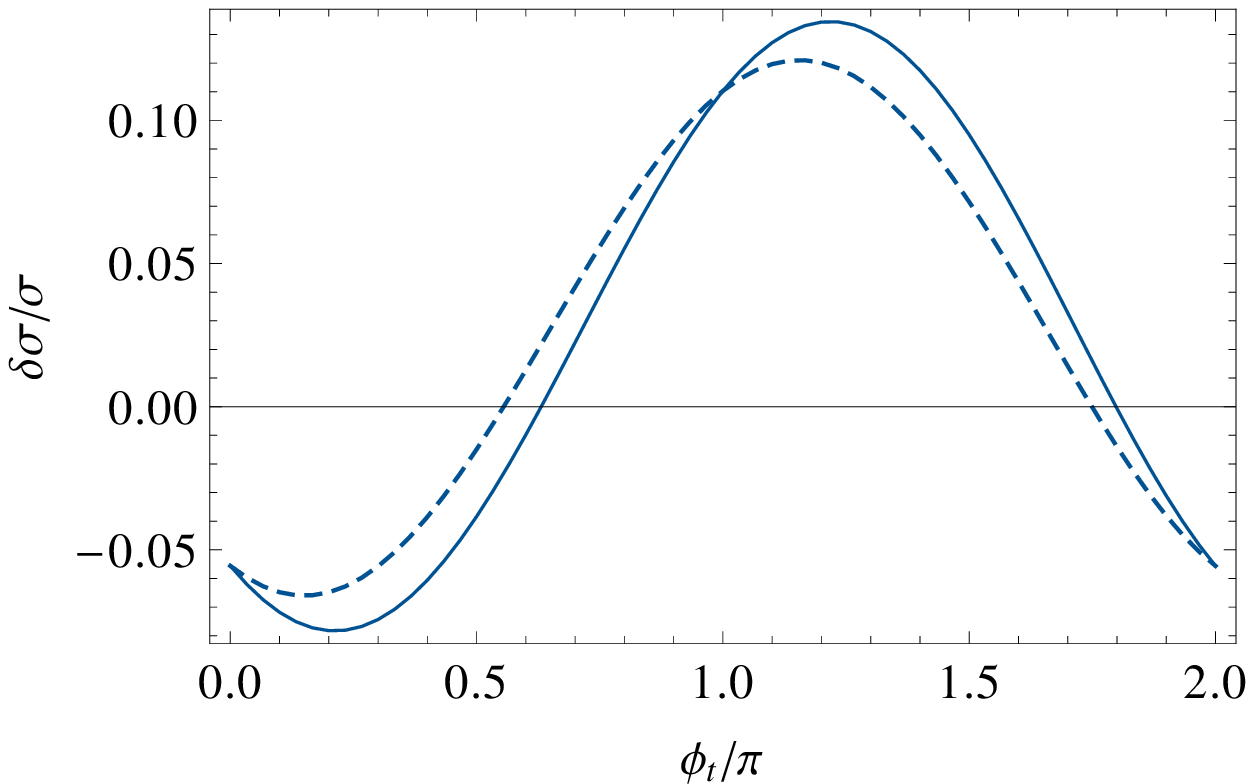}
\includegraphics[scale=.45]{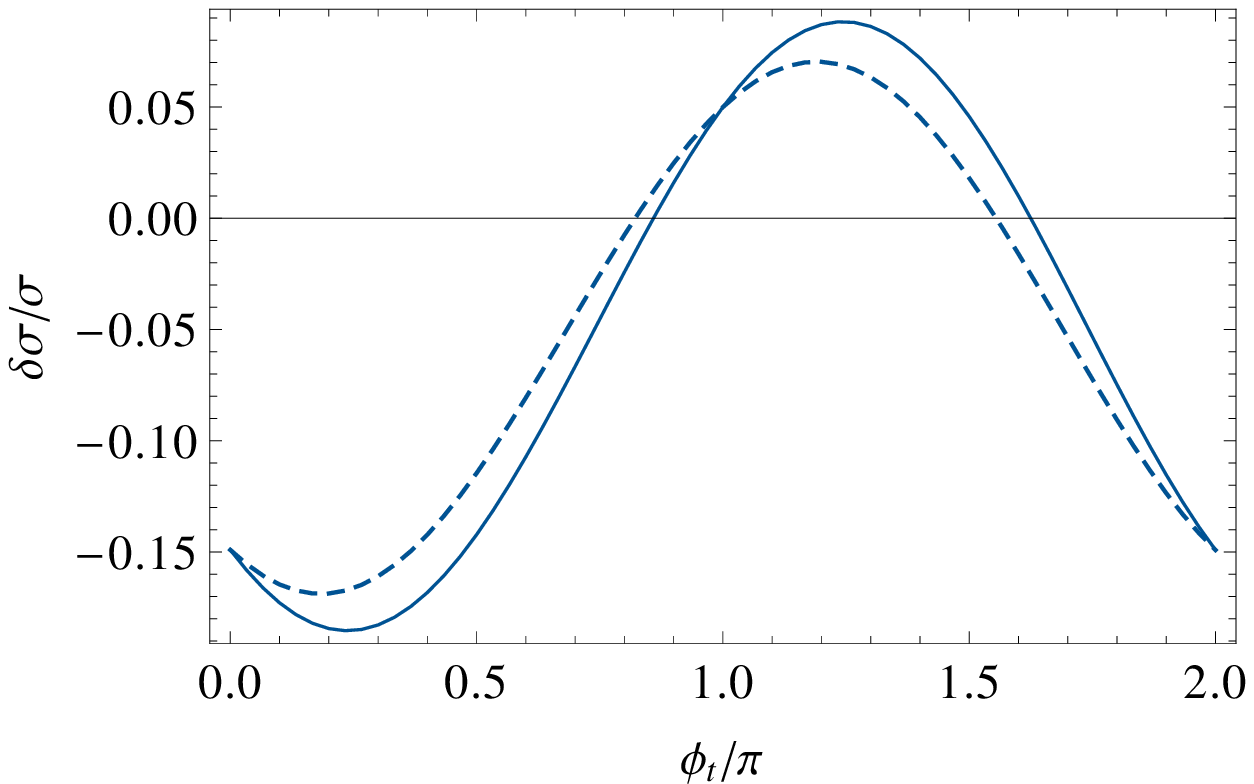}
\caption{The upper plots show $\delta\sigma/\sigma$ for $e^+e^-\to \tilde{\chi}_1^+\tilde{\chi}_2^-$ as a function of the phase $\phi_{t}$ for $M_{\tilde q_{3}}$= 600 (left) and 800 (right) GeV. The lower plots show $\delta\sigma/\sigma$ for $e^+e^-\to
\tilde{\chi}_{1(,L)}^+\tilde{\chi}_{2(,R)}^-$ (left (right)) as a function of $\phi_{A_t}$, for $M_{\tilde q_{3}}$=500 GeV, including/ignoring (solid/dashed) the absorptive parts.}\label{fig:4}
\end{center}
\end{figure}
We study the relative size of the weak 1-loop corrections to the production cross-section, $\displaystyle\delta\sigma/\sigma=(\sigma^{\rm weak}-\sigma^{\rm tree})/\sigma^{\rm tree}$, where $ \sigma^{\rm weak}$ is defined in Ref.\cite{bfmw}, for a $\sqrt{s}=$800 GeV LC.
We find that the only phase that results in effects above  $\mathcal O(\%)$ is $\phi_{A_t}$, and in Fig.~\ref{fig:4} show that the dependence on this phase leads to effects of up to  $\sim 12\%$ for $M_{\tilde q_{3}}$=600 GeV and up to $\sim6\%$ for $M_{\tilde q_{3}}$=800 GeV.
In Fig.~\ref{fig:4} we also show plots of $\delta\sigma/\sigma$ as a functions of
$\phi_{A_t}$ for the case of unpolarised and polarised charginos, showing the impact of ignoring the absorptive parts.
There is seen to be a significant discrepancy between the two results of up to $\sim 2\%$, which could be phenomenologically relevant at linear collider precisions, and therefore for consistency we must ensure the absorptive parts are properly included in calculations in the complex MSSM.

\section{Conclusions}\label{sec:5}

We have defined a consistent renormalisation framework in the on-shell scheme in order to calculate NLO corrections to $e^+e^-\to\tilde\chi^+_i\tilde\chi^-_j$ in the complex MSSM.
In doing this we find that a careful choice of the three on-shell conditions for the chargino and neutralino masses is required, as illustrated in Tab.~\ref{tab:RenormScheme}, and that the phases do not require renormalisation. In addition we use independent field RCs  $\delta Z^{L/R}_{\pm,ij}$ and $\delta\bar{Z}^{R/L}_{\pm,ji}$, in order to include the absorptive parts of loop integrals, and have shown in Fig.~\ref{fig:4} that these absorptive parts in the field RCs can have a 2\% effect on predictions at NLO. Finally, we have calculated the dependence of the cross-section on the phases, showing in Fig.~\ref{fig:4} that the phase $\phi_{A_t}$ has the strongest effect, up to 12\%.

\section{Acknowledgments}

The author gratefully acknowledges support of the DFG through the grant SFB 676, ``Particles, Strings, and the Early Universe'', and thanks her collaborators Alison Fowler, Gudrid Moortgat-Pick and Georg Weiglein.

\begin{footnotesize}


\end{footnotesize}



\begin{thebibliography}{99}
\bibitem{Fowler:2009ay}
  A.~C.~Fowler and G.~Weiglein,
  JHEP {\bf 1001} (2010) 108
  [arXiv:0909.5165 [hep-ph]].

\bibitem{AlisonsThesis}
  A.~C.~Fowler,
  PhD Thesis, 2010, 

\bibitem{bfmw}
A.~Bharucha, A.~Fowler, G.~Moortgat-Pick and G.~Weiglein,
DESY 12-015.

\bibitem{Aad:2011ib}
  G.~Aad {\it et al.} [ ATLAS Collaboration ],
  [arXiv:1109.6572 [hep-ex]].
\bibitem{Chatrchyan:2011zy}
  S.~Chatrchyan {\it et al.} [ CMS Collaboration ],
  [arXiv:1109.2352 [hep-ex]].
\bibitem{Buchmueller:2011sw}
  O.~Buchmueller, R.~Cavanaugh, A.~De Roeck, M.~J.~Dolan, J.~R.~Ellis, H.~Flacher, S.~Heinemeyer, G.~Isidori {\it et al.},
  [arXiv:1110.3568 [hep-ph]].

\bibitem{Desch:2003vw}
  K.~Desch, J.~Kalinowski, G.~A.~Moortgat-Pick, M.~M.~Nojiri and G.~Polesello,
  JHEP {\bf 0402}, 035 (2004)
  [arXiv:hep-ph/0312069].

\bibitem{Rolbiecki:2007se}
  K.~Rolbiecki and J.~Kalinowski,
  Phys.\ Rev.\  D {\bf 76} (2007) 115006
  [arXiv:0709.2994 [hep-ph]].

\bibitem{Eberl:2001eu}
  H.~Eberl, M.~Kincel, W.~Majerotto and Y.~Yamada,
  Phys.\ Rev.\  D {\bf 64} (2001) 115013
  [arXiv:hep-ph/0104109].

\bibitem{Fritzsche:2002bi}
  T.~Fritzsche and W.~Hollik,
  Eur.\ Phys.\ J.\  C {\bf 24} (2002) 619
  [arXiv:hep-ph/0203159].

\bibitem{Chatterjee:2011wc}
  A.~Chatterjee, M.~Drees, S.~Kulkarni and Q.~Xu,
  arXiv:1107.5218 [hep-ph].

\bibitem{Oller:2003ge}
  W.~Oller, H.~Eberl, W.~Majerotto and C.~Weber,
  Eur.\ Phys.\ J.\  C {\bf 29} (2003) 563
  [arXiv:hep-ph/0304006].

\bibitem{Oller:2005xg}
  W.~Oller, H.~Eberl and W.~Majerotto,
  Phys.\ Rev.\  D {\bf 71} (2005) 115002
  [arXiv:hep-ph/0504109].

\bibitem{Fritzsche:2005}
    T.~Fritzsche,
     PhD Thesis, Cuvillier Verlag, G\"{o}ttingen 2005, ISBN 3-86537-577-4.

\bibitem{Kniehl:1996bd}
  B.~A.~Kniehl, A.~Pilaftsis,
  Nucl.\ Phys.\  {\bf B474 } (1996)  286-308.
  [hep-ph/9601390].

\bibitem{Lee:2003nta}
  J.~S.~Lee, A.~Pilaftsis, M.~S.~Carena, S.~Y.~Choi, M.~Drees, J.~R.~Ellis and C.~E.~M.~Wagner,
  Comput.\ Phys.\ Commun.\  {\bf 156} (2004) 283
  [arXiv:hep-ph/0307377].

\end{thebibliography}
\end{document}